\documentclass{article}

\PassOptionsToPackage{numbers, compress}{natbib}

\usepackage[final]{neurips_2023}
\usepackage[utf8]{inputenc}
\usepackage[T1]{fontenc}
\usepackage{xurl}
\usepackage{hyperref}
\usepackage{enumitem}
\usepackage{booktabs}
\usepackage{amsfonts}
\usepackage{nicefrac}
\usepackage{microtype}
\usepackage{xcolor}
\usepackage{graphicx}
\usepackage{tabularx}
\usepackage{xcolor}
\usepackage{cleveref}
\usepackage{caption} 
\title{A Grading Rubric for AI Safety Frameworks}

\author{
Jide Alaga\thanks{Corresponding author: \href{mailto:jide.alaga@governance.ai}{jide.alaga@governance.ai}.} \quad
Jonas Schuett \quad
Markus Anderljung \\ \\
Centre for the Governance of AI
}

\begin{document}

\maketitle

\begin{abstract}
Over the past year, artificial intelligence (AI) companies have been increasingly adopting AI safety frameworks. These frameworks outline how companies intend to keep the potential risks associated with developing and deploying frontier AI systems to an acceptable level. Major players like Anthropic, OpenAI, and Google DeepMind have already published their frameworks, while another 13 companies have signaled their intent to release similar frameworks by February 2025. Given their central role in AI companies' efforts to identify and address unacceptable risks from their systems, AI safety frameworks warrant significant scrutiny. To enable governments, academia, and civil society to pass judgment on these frameworks, this paper proposes a grading rubric. The rubric consists of seven evaluation criteria and 21 indicators that concretize the criteria. Each criterion can be graded on a scale from A (gold standard) to F (substandard). The paper also suggests three methods for applying the rubric: surveys, Delphi studies, and audits. The purpose of the grading rubric is to enable nuanced comparisons between frameworks, identify potential areas of improvement, and promote a race to the top in responsible AI
development.
\end{abstract}

\begin{table}[ht]
    \includegraphics[width=\linewidth]{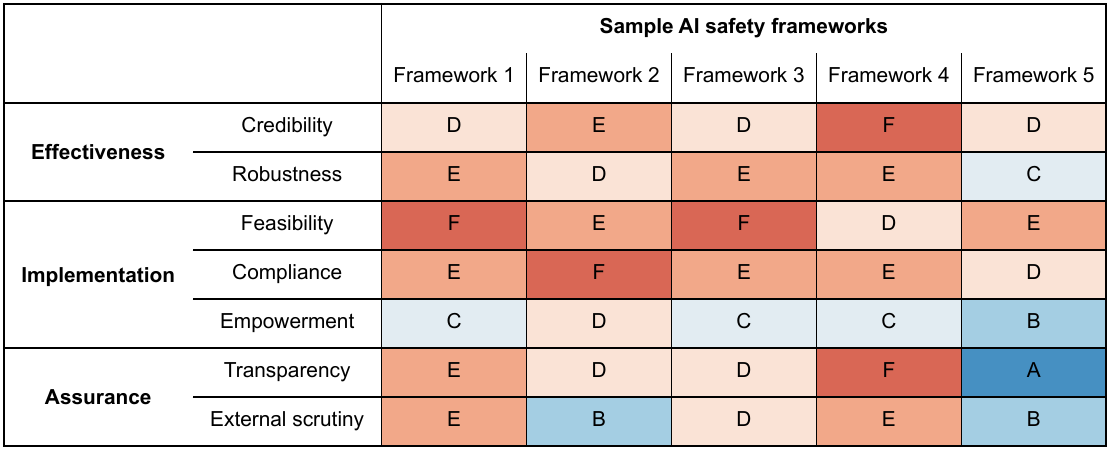}
    \caption{Sample grading to illustrate the evaluation criteria and quality tiers}
    \label{tab:3-sample-grading}
\end{table}
\newpage

\section*{Executive summary}\label{executive-summary}

This paper proposes a grading rubric for passing judgment on AI safety frameworks. It is divided into four sections. First, we explain what AI safety frameworks are and discuss why passing judgment on them is important. Next, we propose seven evaluation criteria and 21 indicators to concretize the criteria, along with a six-tiered grading system. We also suggest three ways in which different actors can use the rubric to grade existing frameworks. Finally, we discuss potential limitations of the proposed rubric.

\textbf{What are AI safety frameworks?} AI safety frameworks are risk management policies intended to keep the potential risks associated with developing and deploying frontier AI systems to an acceptable level. These frameworks typically focus on catastrophic risks (e.g. from the use of chemical or biological weapons, cyberattacks, or loss of control). They specify, among other things: (1) how developers analyze the potential ways in which AI systems could lead to catastrophic outcomes, (2) how they gather evidence about a system's capabilities, (3) what safety measures would be adequate for a given level of capabilities, and (4) how developers intend to ensure that they adhere to the framework and maintain its effectiveness (\Cref{what-are-ai-safety-frameworks}).

\textbf{The case for passing judgment on AI safety frameworks.} There are at least five main reasons why grading AI safety frameworks might be valuable. (1) The grading process might identify shortcomings of safety frameworks, and thereby ways they can be improved. (2) Passing judgment on frameworks might incentivise companies to improve their frameworks in response to poor grades, or from a desire to be seen as a leader in responsible AI -- especially if the grades are public and viewed as legitimate. This dynamic could ultimately lead to a race to the top in safety. (3) In the future, safety frameworks might be folded into regulation, and regulators might be required to make assessments about their adequacy. Since developing the necessary assessment skills and tools will likely take time, regulators and other third parties may want to start today. (4) Passing judgements on AI safety frameworks could inform the public discussion about companies' commitments to safety. (5) Developing an evaluation framework for AI safety frameworks can contribute to the development of industry best practices (\Cref{the-case-for-passing-judgment-on-ai-safety-frameworks}).

\textbf{Grading rubric.} To enable governments, researchers, and civil society to pass judgment on AI safety frameworks, we propose a new grading rubric. The rubric consists of seven evaluation criteria divided into three categories. We also propose 21 corresponding indicators that concretize the criteria. \autoref{tab:1-Overview} provides an overview of the categories and evaluation criteria, while \autoref{tab:4-evaluation-criteria} also includes the indicators (\Cref{grading-rubric-for-ai-safety-frameworks}).

\begin{table}[ht!]
    \centering
    \includegraphics[width=\linewidth]{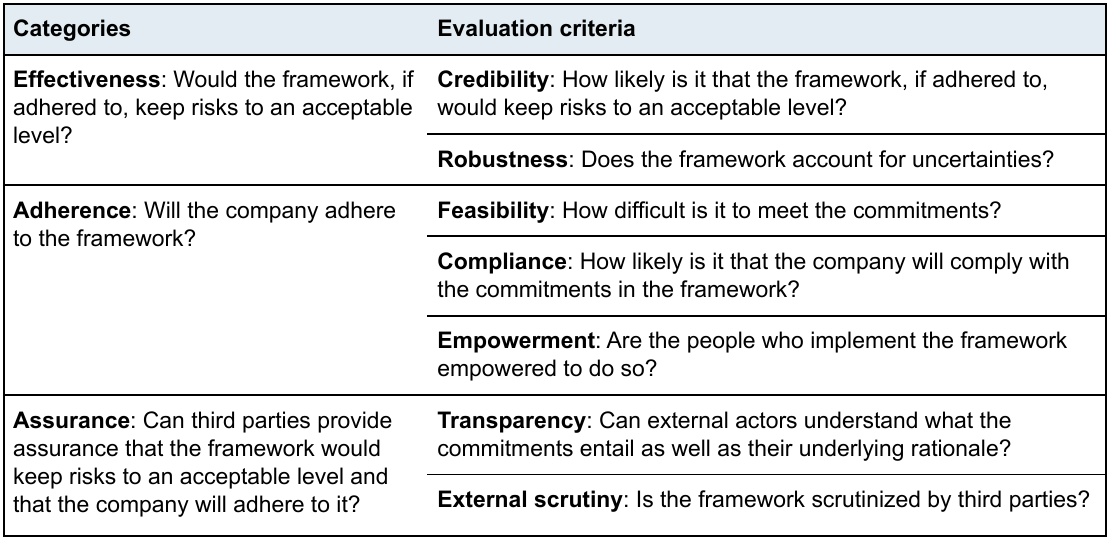}
    \caption{Overview of the evaluation criteria}
    \label{tab:1-Overview}
\end{table}
\newpage

\textbf{Quality tiers.} The evaluation criteria can be graded on a scale from A (gold standard) to F (substandard). The tiers are defined in terms of (1) how much the frameworks satisfy the specified evaluation criteria, (2) how much room for improvement they leave, and (3) to what extent the demonstrated level of effort is commensurate with the stakes. \autoref{tab:2-quality-tiers} contains a description of the six quality tiers (\Cref{quality-tiers}).

\begin{table}[ht!]
    \includegraphics[width=\linewidth]{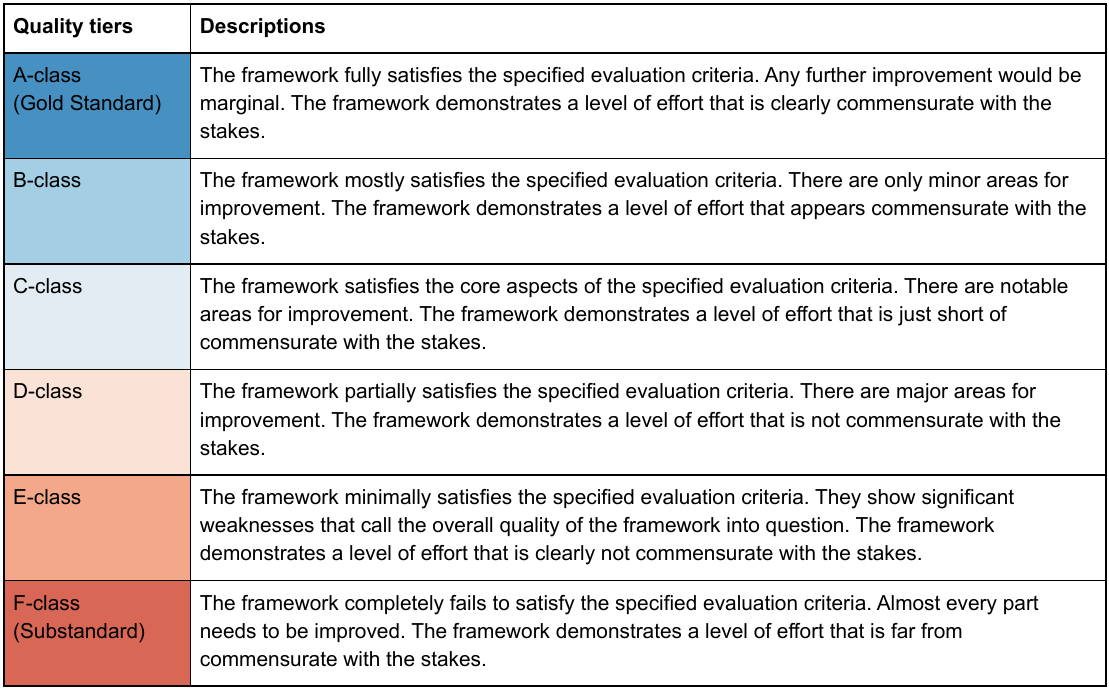}
    \caption{Description of quality tiers}
    \label{tab:2-quality-tiers}
\end{table}

\textbf{Applying the evaluation criteria.} We also suggest three ways in which different actors can use the rubric to grade AI safety frameworks. (1) Surveys, which are filled out by external researchers to ensure high degrees of independence. (2) Delphi studies, which combine surveys and workshops. They allow participants to update their grades after a workshop in which participants discuss the rationales behind their grades. (3) Audits, which are conducted by a third party who is given access to confidential information and key personnel (\Cref{how-to-apply-the-grading-rubric}).

\textbf{Limitations.} Our grading rubric has at least six limitations. (1) The evaluation criteria do not offer specific guidance on what commitments would be desirable (e.g. they do not contain any sample commitments). As such, evaluation results might not translate into actionable recommendations. (2) Many evaluation criteria are difficult to measure objectively, which might lead to inconsistent scores. (3) Given the above, assessing the criteria involves a number of judgment calls, which could limiting the number of people who could credibly assess the quality of frameworks. (4) The evaluation criteria are unlikely to be exhaustive. (5) It might be difficult to differentiate the six quality tiers. (6) The grading rubric does not weigh the evaluation criteria differently, despite the fact that they are unlikely to be equally important. We look forward to seeing others use, build on, and critique our rubric (\Cref{limitations}).
\newpage

\section{Introduction}\label{introduction}

\subsection{What are AI safety frameworks?}\label{what-are-ai-safety-frameworks}

AI safety frameworks are risk management policies which outline how AI companies intend to keep the potential risks associated with developing and deploying frontier AI systems\footnote{By ``frontier AI systems'', we mean highly capable general-purpose AI models that can perform a wide variety of tasks and match or exceed the capabilities present in the most advanced models \citep{dsit2024}.} to an acceptable level. These frameworks primarily aim to address AI-enabled catastrophic risks (e.g. from the use of chemical or biological weapons, cyberattacks, or loss of control), though they could also address other risks that might be considered unacceptable.\footnote{By ``catastrophic risks'', we mean risks that can result in $\geq$10,000 human deaths or $\geq$\$100 billion in economic costs to society. This is in line with definitions used by others \citep{anthropic2023, shevlane2023}. However, it is worth noting that this is a somewhat arbitrary threshold.} Since there is currently no universal standard for what constitutes an ``acceptable'' level of risk, each developer determines this threshold themselves. Safety frameworks typically have four main components:\footnote{These four components are also present in other risk management frameworks, such as ISO/IEC 23894 \citep{iso2023} or ISO 31000 \citep{iso2018}. For a different breakdown of components, see \citep{metr2024}.}

\textbf{Risk identification.} Companies outline how they analyze the potential ways in which AI systems could lead to catastrophic outcomes. This typically involves detailing threat models and explicitly specifying the threat vectors and risk scenarios that concern them most within their safety frameworks \citep{metr2024}. These threats are often framed in terms of dangerous capabilities that the AI system could possess, which might enable it to cause significant harm either through intentional misuse or unintended accidents \citep{shevlane2023, phuong2024}. It also involves setting risk thresholds for each category of threat, and particularly specifying the threshold at which their systems would pose unacceptable risks \citep{koessler2024}.

\textbf{Risk assessment.} Companies outline how they gather evidence about a system's capabilities. This process typically involves specifying a set of model evaluations designed to detect dangerous capabilities related to the identified threats \citep{phuong2024, shevlane2023}. Developers also establish clear criteria for interpreting the evaluation results, setting thresholds that indicate when a system possesses these dangerous capabilities. Often, attempts are made to assess what could be achieved using the AI system that they could not without it, though efforts to tie model evaluations to such real-world capabilities are still nascent.

\textbf{Risk mitigation.} Companies outline the safety measures that would be adequate for a given level of capabilities. To do this, developers specify safeguards to be implemented at each risk level for every identified threat category. They also commit to implementing these safeguards once model evaluation results indicate the corresponding risk level of the system. Safety protocols also usually include ``red lines'', which are implemented whenever model evaluations suggest that a system is approaching an unacceptable level of risk.\footnote{OpenAI’s Preparedness Framework can also be said to do this \citep{openai2023}. It requires that if a deployed model’s capabilities are above ``Medium'', mitigations should be put in place to take risk back down to ``Medium''.} These are thresholds that indicate when the risk is too high for developers to continue scaling or deploying a system. When a system crosses a ``red line'', developers must pause AI progress and reassess how to make the system safe before proceeding further \citep{alaga2023}.

\textbf{Risk governance.} Companies outline how they intend to ensure that they adhere to the framework and maintain its effectiveness. This may include commitments to red team model evaluations, keep external stakeholders informed about evaluation results \citep{kolt2024a}, monitor compliance with safety frameworks, and implement oversight and governance procedures. These procedural commitments help to ensure that the safety framework remains a top priority and is consistently applied throughout the development process.

Although this is what existing safety frameworks typically look like, in principle, they can vary significantly. For example, risk assessments could be expanded to consider not only the inherent capabilities of a system, but also information about the external world in which the system will be deployed. This could involve quantitative risk thresholds that take into account both the likelihood and potential severity of identified threats. Additionally, safety frameworks can establish red lines beyond decisions about whether to deploy or train systems. Developers could establish additional red lines that guide decisions on whether to open-source systems, for example. The flexibility over what constitutes a safety framework allows them to be tailored to different contexts and the needs of different developers while still mitigating catastrophic risks.

\subsection{The case for passing judgment on AI safety frameworks}\label{the-case-for-passing-judgment-on-ai-safety-frameworks}

Below, we discuss four arguments for why passing judgment on AI safety frameworks is important.

\textbf{Identifying shortcomings.} Since it is extremely difficult to keep risks from the development and deployment of (especially future) frontier AI systems to an acceptable level with high confidence, we should expect that safety frameworks will have significant shortcomings. Initial frameworks are likely to need significant improvement over time, as AI systems advance and pose greater risks. The grading process might identify such shortcomings, which allows companies to improve their frameworks. This is analogous to review processes in other domains (e.g. the peer review~process in science).

\textbf{Incentivising a race to the top.} Many AI companies want to be perceived as responsible actors, and as a result they might try to improve their frameworks in response to poor grades,or from a desire to be viewed as ``best in class'' -- especially if the grades are public and the grading is viewed as legitimate. This dynamic could ultimately lead to a ``race to the top'' in safety standards, where companies strive to demonstrate the most comprehensive and effective safety frameworks, leading to an overall increase in the quality of safety standards across the industry.

\textbf{Preparing for regulation.} In the future, there may be regulatory requirements to implement AI safety frameworks. In that case, the ability to pass judgment on safety frameworks may become a central responsibility for regulators, making it essential to develop these evaluation skills now. This holds under different regulatory approaches.\footnote{For more information on regulatory approaches for frontier AI, see \citep{schuett2024}.} Under a rules-based approach, companies may be required to implement a specific safety framework, and regulators would need to evaluate them to assess compliance. Under a goals-based approach, companies may be required to keep risks from frontier AI systems to an acceptable level, without specifying how to do that, and regulators would need to evaluate the extent to which a safety framework achieves this goal.

\textbf{Informing the public.} Passing judgements on AI safety frameworks could inform the public discussion about a company's commitments to safety. When external actors assess the safety frameworks of different AI companies and make their findings public, it helps the public gauge the reliability of these frameworks. This external validation is particularly important because of the complex and technical nature of AI systems, which most members of the public lack the expertise to evaluate independently. Moreover, companies may engage in ``safety washing'', claiming their frameworks are better than they actually are, and without third-party verification, the public may struggle to discern the truth.

\subsection{Related work}\label{related-work}

Given that AI safety frameworks have only recently emerged, scholarship on the topic is scarce. Existing work can be broadly categorized into four main areas:

\textbf{Existing safety frameworks.} To date, only four companies have published AI safety frameworks as defined above. Anthropic published their Responsible Scaling Policy (RSP) in September 2023 \citep{anthropic2023}. A few months later, they also shared initial reflections from the implementation of their framework \citep{anthropic2024}. OpenAI published their Preparedness Framework (Beta) in December 2023 \citep{openai2023}, Google DeepMind their Frontier Safety Framework in May 2024 \citep{googledeepmind2024}, and Magic their AGI Readiness Policy in July 2024 \citep{magic2024}. Another 13 companies signed the Frontier AI Safety Commitments \citep{dsit2024} at the AI Seoul Summit in May 2024, committing to producing safety frameworks ahead of the Paris AI Action Summit in February 2025.

\textbf{Recommendations for safety frameworks.} Several scholars and practitioners have also made recommendation for safety frameworks. METR, an organization which played a crucial role in popularizing the concept of AI safety frameworks, outlined five main components that such frameworks should include \citep{metr2023, metr2023a}. They also reviewed common elements of existing safety frameworks \citep{metr2024}. Similarly, the UK Department for Science, Innovation and Technology (DSIT) proposed seven practices to be incorporated into responsible capability scaling policies \citep{dsit2023}. More recently, the Frontier AI Safety Commitments list several elements that should be included in safety frameworks \citep{dsit2024}.

\textbf{Reviews of existing safety frameworks.} A few scholars have already conducted reviews of existing frameworks. Anderson-Samways et al. evaluated Anthropic's RSP against DSIT's guidance on responsible capability scaling \citep{anderson-samways2024}. Similarly, \'{O} h\'{E}igeartaigh et al. conducted a rapid review of company statements on AI safety frameworks ahead of the 2023 AI Safety Summit in Bletchley Park \citep{oheigeartaigh2024}. Each statement was evaluated against 42 safety practices proposed by \citep{dsit2023}. Additionally, SaferAI compared Anthropic's RSP against OpenAI's Preparedness Framework \citep{saferai2024}.

\textbf{Evaluation criteria.} There is only a single source that proposes criteria for AI safety frameworks. Titus proposed nine criteria that safety frameworks should satisfy to robustly address the risks associated with building advanced AI systems \citep{titus2024}. While Titus' work is similar in intent to the present paper, our approach aims to go a step further by presenting the criteria as tools for evaluation. This paper not only identifies components that indicate a safety framework's likelihood of success, but also provides a rubric for differentiating between various levels of quality.

\section{Grading rubric for AI safety frameworks}\label{grading-rubric-for-ai-safety-frameworks}

In this section, we propose a grading rubric for AI safety frameworks. The rubric distinguishes between three categories of evaluation criteria: effectiveness (\Cref{effectiveness}), adherence (\Cref{adherence}), and assurance (\Cref{assurance}). They correspond roughly to the three outcomes described in the Frontier AI Safety commitments \citep{dsit2024}.\footnote{The outcomes are: (1) Organisations effectively identify, assess and manage risks when developing and deploying their frontier AI models and systems. (2) Organisations are accountable for safely developing and deploying their frontier AI models and systems. (3) Organisations' approaches to frontier AI safety are appropriately transparent to external actors, including governments.} Within each category, we define evaluation criteria and indicators that concretize the criteria. \autoref{tab:4-evaluation-criteria} provides an overview of all criteria and indicators. Each criterion can be graded on a scale from A (gold standard) to F (substandard) (\Cref{quality-tiers}).

\begin{table}[ht!]
    \centering
    \includegraphics[width=\linewidth]{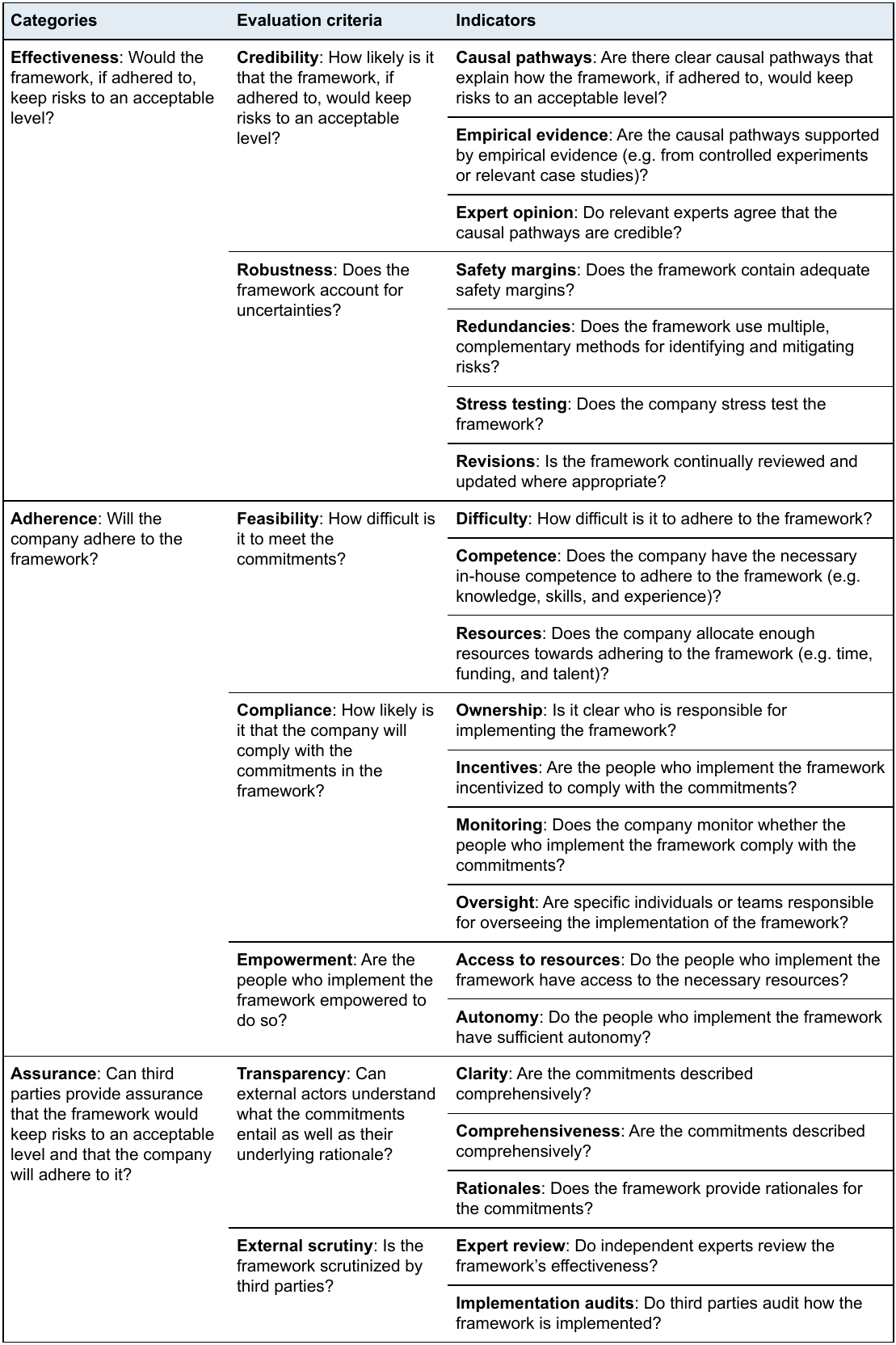}
    \caption{Overview of the evaluation criteria and indicators}
    \label{tab:4-evaluation-criteria}
\end{table}

\subsection{Effectiveness}\label{effectiveness}

First, we propose two criteria that can be used to evaluate a framework's effectiveness, i.e. the degree to which the framework would, if adhered to, keep risks to an acceptable level. The two criteria are credibility (\Cref{credibility}) and robustness (\Cref{robustness}).

\subsubsection{Credibility}\label{credibility}

The first effectiveness-related criterion evaluates how likely it is that the framework, if adhered to, would keep risks to an acceptable level. It focuses on the evidence supporting key decisions in safety frameworks, such as the choice of threat models, risk thresholds, model evaluations, and safeguards.

Scoring highly on this criterion means that evaluators find these decisions justifiable. Specifically, it means that there are strong reasons to believe that these decisions are not only relevant to catastrophic risk reduction, but also have a high likelihood of effectively achieving their intended outcomes. Note that this is an objective measure, and not necessarily linked to existing best practices. For example, it is possible that in the spectrum of potential safety frameworks, the ideal framework would lean more toward not training any frontier models until an objective, formal, verification of safety can be established, rather than one where companies simply adhere to all currently recommended best practices for safety and risk management.

Why is this criteria important for making assessments on safety frameworks? In short, the stronger the evidence base behind a safety framework, the more confident we should be that it would appropriately identify and address risks. Safety frameworks with a weak or insufficient evidence base may include commitments that are ineffective, misdirected, or even counterproductive in mitigating catastrophic risks. This can create a false sense of security, causing actors to believe that the policy is more reliable than it actually is. To concretize this criterion, graders could use the following three indicators:

\textbf{Causal pathways.} Are there clear causal pathways that explain how the framework, if adhered to, would keep risks to an acceptable level? These causal pathways should link the components of the safety framework (threat models, risk thresholds, model evaluations, etc.) to its main objective (keeping risks to an acceptable level). The pathways should be logically consistent and based on reasonable assumptions.

\textbf{Empirical evidence.} Are the causal pathways supported by empirical evidence? Evidence might stem from controlled experiments or relevant case studies. In general, evidence is stronger if it is scientifically validated (e.g. if it has been peer-reviewed and successfully replicated). However, due to the novelty of AI safety frameworks (and the slowness of the academic publication process), scientific evidence will often not yet exist and evidence from other domains might not generalize to an AI context.

\textbf{Expert opinion.} Do relevant experts agree that the causal pathways are credible? By ``relevant experts'', we mean scholars and practitioners with AI safety or governance expertise. A high level of agreement would suggest that a safety framework represents the state-of-the-art. However, it might be difficult to reach expert consensus.\footnote{However, a recent expert survey found a remarkably high level of agreement with various statements about AI safety and governance measures \citep{schuett2023a}.} And even if consensus is reached, it would only provide weak evidence for the effectiveness of a safety framework (e.g. because experts might be biased). Against this background, expert opinion should be seen as a weaker indicator than empirical evidence.

\subsubsection{Robustness}\label{robustness}

The second effectiveness-related criterion evaluates the framework's robustness, i.e. the extent to which it accounts for potential changes in the risk landscape and potential failures of risk assessment and mitigation measures. Robustness is important because many risks from AI are poorly understood and rapidly evolving. Companies might therefore fail to foresee some risk events. Such events can be conceptualized as ``black swans'' \citep{aven2013, aven2016, kolt2024, taleb2010} or ``unknown unknowns'' \citep{gilbert2016}. At the same time, many risk assessment and mitigation measures are nascent and best practices do not yet exist. As a result, some threat models might be incorrect, some model evaluations might fail to identify dangerous model capabilities, and some mitigation measures might be less effective than expected. We wish to emphasize that robustness comes at a cost. It is possible that measures intended to increase robustness will ultimately be incommensurate to the gain in reduced risk. Graders should take this possibility into account when assessing robustness. To concretize the robustness criterion, graders could use the following four indicators:

\textbf{Safety margins.} Does the safety framework contain adequate safety margins? For example, companies might commit to implement safety measures before they are necessary (e.g. before certain capability thresholds are reached) or they could implement stronger measures than necessary (e.g. measures that would be adequate for higher capability levels).

\textbf{Redundancies.} Does the safety framework use multiple, complementary methods for assessing and mitigating risks? Redundancies make the framework more robust against failures of individual measures. This approach, which is often referred to as ``defense in depth'' \citep{cotton-barratt2020, holmberg2017, perrow2000} or the ``Swiss cheese model'' \citep{larouzee2020, reason1990, reason1990a}, is very common in other safety-critical domains like cyber security \citep{ee2024, stytz2004}, nuclear energy \citep{fleming2002}, and aviation \citep{li2011}.

\textbf{Stress testing.} Does the company stress test the safety framework? For example, they could try to anticipate and plan for potential failure modes (``pre-mortems''). They could create a list of worst-case scenarios and evaluate whether the framework would provide adequate protection against each of them. In the nuclear industry, this approach has been referred to as ``deterministic risk assessments'' \citep{kirchsteiger1999, richardson1996}. The company could also engage independent third parties to red team the framework (e.g. challenging key assumptions or trying to identify potential weaknesses).

\textbf{Revisions.} Is the safety framework continually reviewed and updated where appropriate? Frameworks should reflect the state of the art and industry best practices. They should ideally be treated as living documents, being regularly reassessed and revised to incorporate new scientific findings and lessons learned from past implementations. Ideally safety frameworks would also state the conditions under which they are valid, and specify concrete points where the framework is to be updated before development or deployment proceeds \citep{anthropic2023, googledeepmind2024, magic2024}.

\subsection{Adherence}\label{adherence}

Next, we propose three criteria that can be used to evaluate the extent to which companies will adhere to their AI safety frameworks: feasibility (\Cref{feasibility}), compliance (\Cref{compliance}), and empowerment (\Cref{empowerment}).

\subsubsection{Feasibility}\label{feasibility}

The first adherence-related criterion evaluates how difficult it is to meet the commitments. In essence, it seeks to answer the question: are the proposed safety measures realistic, or are they overly ambitious given the developer\textquotesingle s current capabilities and constraints? The value of satisfying this criteria is that it avoids scenarios where developers are heavily dependent on these commitments for risk mitigation, only to discover that putting them in place is not actually possible. To concretize this criterion, graders could use the following three indicators:

\textbf{Commitment difficulty.} What is the inherent difficulty of implementing the proposed measures? The following questions could be used to gather evidence about the difficulty of different activities under the framework: Does the science and technology to meet the commitments exist yet? Will it exist by the time the commitments would require it to? Have other actors successfully done what the framework requires?

\textbf{Developer competence.} Does the company have the necessary in-house competence to adhere to the framework? The people who implement the framework need certain knowledge, skills, and experience. A track record of performing similar activities in the past suggests that the company has the necessary competence. Companies may compensate for a lack of competence by partnering with individuals or organizations. Graders should also take into account any steps the company might have taken to build up competence training or hiring.

\textbf{Resources committed.} Does the company allocate enough resources towards adhering to the framework? Among other things, companies need time, funding, and talent. Setting unrealistic deadlines, lacking sufficient funds, or hiring people with insufficient skills can each undermine the feasibility of the commitments. If the commitments involve uncommon collaborations or favorable regulatory treatment, the company also needs social and political capital.

\hypertarget{compliance}{%
\subsubsection{Compliance}\label{compliance}}

The second adherence-related criterion evaluates how likely it is that the company will comply with the commitments in the safety framework. More concretely, it evaluates the extent to which developers take proactive measures to ensure that the people who implement the framework comply with the commitments as intended. Satisfying this criteria well should therefore provide evaluators with confidence that personnel will be motivated to follow these policies, even if they are viewed as inconvenient chores. To concretize this criterion, graders could use the following four indicators:

\textbf{Ownership.} Is it clear who is responsible for implementing the safety framework? Implementing the framework will likely involve different people from different teams with different responsibilities \citep{robinson2024, schuett2023}. If these responsibilities are not clearly assigned and coordinated, gaps in risk coverage can occur \citep{bantleon2021}. To this end, companies might implement the Three Lines Model \citep{davies2018, instituteofinternalauditors2013, instituteofinternalauditors2020}, Combined Assurance Framework \citep{decaux2015, huibers2015}, or other risk governance frameworks.

\textbf{Incentives.} Are the people who implement the framework incentivized to comply with the commitments? Companies should ideally use a combination of positive and negative incentives (``carrots and sticks''). Positive incentives might include rewards and recognition for high degrees of compliance. Negative incentives might include disciplinary action and financial consequences for non-compliance.

\textbf{Monitoring.} Does the company monitor whether commitments are adhered to as intended? Companies may track whether certain performance targets are met, conduct regular drills, and establish whistleblowing schemes with strong protections for reporting non-compliance. The presence of measures designed to track how often and how well commitments are being adhered to are necessary factors for ensuring compliance.

\textbf{Oversight.} Are specific individuals or teams responsible for overseeing the implementation of the framework? Without such oversight, accountability gaps can occur, where no one is held accountable for implementation failures. To avoid conflicts of interest, the people who oversee the implementation of the framework should ideally be separate from the people who implement the framework.

\hypertarget{empowerment}{%
\subsubsection{Empowerment}\label{empowerment}}

The third adherence-related criterion evaluates the extent to which the people who implement the safety framework are empowered to do so. It assesses the extent to which safety frameworks include measures to protect personnel from factors that might undermine their efforts. If they do, then developers should be able to avoid scenarios where Safety Policies are theoretically sound and employees make sincere efforts to adhere to them, but they still fail because personnel aren't sufficiently equipped to properly fulfill them. To concretize this criterion, graders could use the following two indicators:

\textbf{Access to resources.} Do the people who implement the framework have access to the necessary resources? This might include time, training, funding, information, and social and political capital, among other things. It might also include ensuring that the ease of access to such resources is sufficient, as excessive bureaucracy and red tape can disrupt safety efforts by slowing down processes when swift action is required.

\textbf{Autonomy.} Do the people who implement the framework have sufficient autonomy? In particular, are they protected against interference from other actors with competing interests (e.g. the product team)? Without sufficient autonomy, even well-resourced individuals and teams might fail to implement the framework.

\hypertarget{assurance}{%
\subsection{Assurance}\label{assurance}}

Finally, we propose two criteria that can be used to evaluate the extent to which third parties can provide assurance that safety frameworks would keep risks to an acceptable level and that companies will adhere to them: transparency (\Cref{transparency}) and external scrutiny (\Cref{external-scrutiny}).

\hypertarget{transparency}{%
\subsubsection{Transparency}\label{transparency}}

The first assurance-related criterion evaluates how accessible and well-communicated the commitments are. Do they contain the information necessary to scrutinize and grade them (see \Cref{external-scrutiny})? If the commitments in a safety framework are imprecise or omit crucial information, it creates the possibility for two individuals to read them and walk away with different interpretations of what a developer has committed to do, and how they plan to do it. This ambiguity can make it difficult to assess the framework's potential effectiveness, especially if the criteria for success or failure are also unclear. Therefore it is important for developers to be penalized for being insufficiently transparent. To concretize the transparency criterion, graders could use the following three indicators:

\textbf{Clarity.} Are the commitments described clearly? To avoid misinterpretations, the language used to describe the commitments should be understandable (i.e. avoid jargon and technical terms), precise (i.e. avoid ambiguous terms), concise (i.e. be as short as possible), and consistent (i.e. use the same terms and concepts throughout the document).

\textbf{Comprehensiveness.} Are the commitments described comprehensively? They should not leave out information about key elements of the framework (e.g. about the risk identification process or pass/fail criteria for model evaluations). The company should ideally also publish results from any audits or expert reviews. The descriptions should be comprehensive enough for third parties to implement the safety framework independently. This is similar to replicability in science, where the methods must be described with sufficient detail to allow other scientists to replicate a study.

\textbf{Rationales.} Does the framework provide rationales for the commitments? The framework should contain or link to explanations and justifications of key design choices (e.g. why they think that certain safety measures would be adequate for a given level of capabilities). Providing the reasoning behind these decisions is particularly important for assessing the framework's credibility (\Cref{credibility}) and feasibility (\Cref{feasibility}).

\hypertarget{external-scrutiny}{%
\subsubsection{External scrutiny}\label{external-scrutiny}}

The second assurance-related criterion evaluates the extent to which the safety framework includes explicit commitments to be periodically scrutinized by third parties. Such external scrutiny is important both to provide stakeholders with reliable information about both the effectiveness of the framework and the extent to which it is being adhered to. This is important even where companies are whole-heartedly attempting to develop and implement a high-quality framework, as they may well have important blind spots and to give external stakeholders a reason to trust their efforts. To concretize this criterion, graders could use the following two indicators:

\textbf{Expert review.} Do independent experts review the framework's effectiveness? Such reviews provide additional assurance that the framework would, if adhered to, keep risks to an acceptable level (see
\Cref{effectiveness}). Independent experts may include researchers or policy advisors from academic institutions, think tanks, or civil society organizations. They may identify potential blindspots, false assumptions, or reasoning errors. By identifying potential shortcomings, reviews help to ensure that the framework is theoretically sound and reflects current best practices in AI safety and governance. They should be conducted during the initial drafting phase of the framework and whenever the framework is updated.

\textbf{Implementation audits.} Do third parties audit how the framework is implemented? Such audits provide additional assurance that the company has implemented the framework as intended (see \Cref{adherence}). Implementation audits could be conceptualized as a type of ``governance audits'' \citep{mokander2023}. They can help to overcome information asymmetries between the company and external stakeholders \citep{cihon2021}: it is difficult for people outside the company to know whether the company has actually implemented their framework. They should be conducted on a regular basis (e.g. annually).

\hypertarget{quality-tiers}{%
\subsection{Quality tiers}\label{quality-tiers}}

Each evaluation criterion can be graded on a scale from A (gold standard) to F (substandard). The six quality tiers are defined in terms of (1) the extent to which the frameworks satisfy the specified evaluation criteria (ranging from ``fully satisfy'' to ``completely fail''), (2) how much room for improvement they leave (ranging from ``any further improvement would be marginal'' to ``almost every part needs to be improved''), and (3) the extent to which the demonstrated level of effort is commensurate with the stakes of catastrophic AI risk mitigation (ranging from ``clearly commensurate'' to ``far from commensurate''). \autoref{tab:2-quality-tiers} contains a description of the six quality tiers.

\hypertarget{how-to-apply-the-grading-rubric}{%
\section{How to apply the grading rubric}\label{how-to-apply-the-grading-rubric}}

In this section, we suggest three methods for applying the grading rubric: surveys (\Cref{survey}), Delphi studies (\Cref{delphi-study}), and audits (\Cref{audit}). We think it makes most sense to apply the grading rubric to the evaluation criteria, grading each on a scale from A to F. \autoref{tab:3-sample-grading} illustrates this for five hypothetical frameworks. Although it would also be possible to grade each of the 21 indicators, this will often be too time-intensive. Inversely, it would also be possible to only provide a single overall score for each framework or to grade each of the three categories, but we believe this would lose important nuances.

\subsection{Survey}\label{survey}

One method for applying the grading rubric is to conduct a survey. This approach would involve three main steps:

\textbf{Survey design.} For each safety framework, the survey would ask participants to evaluate each criterion on a scale from A to F. The questions would briefly describe the criteria and list the corresponding indicators. Participants may also provide rationales for their responses, state key uncertainties, and potentially also suggest ways to improve the framework.

\textbf{Sample.} The survey could be sent to independent AI safety and governance experts from governments, academia, and/or civil society. In general, a larger sample size is preferable (e.g. \textgreater20 participants). However, since relevant expertise is scarce and filling out the survey is fairly time-intensive, it will often not be possible to reach a very large sample size.

\textbf{Analysis.} After the survey, the responses could be aggregated and the average grade for each criterion can be reported. If the results are written up in a report, then the variance, rationales, and main uncertainties for each criterion can also be documented. These insights can be valuable for identifying potential gaps or ambiguities in the frameworks.

A key advantage of this approach is that it is likely to be less resource intensive than the Delphi study. Further, it provides a clear and interpretable output (a specific grade for each criteria), while still being able to draw on the difficult-to-articulate expertise and judgment from the evaluators.

However, it is worth noting that a highly structured questionnaire may not capture all the nuances and context-specific factors that could influence the effectiveness of an AI safety framework. Some graders may find the rigid format of the survey limiting, as it does not allow for the same level of in-depth discussion and exploration as other methods.

\hypertarget{delphi-study}{%
\subsection{Delphi study}\label{delphi-study}}

Another method for applying the grading rubric is through Delphi studies \citep{cooke2004, hsu2007}. There are three main steps in this process. First, participants fill out a survey in which they are asked to evaluate each criterion and provide rationales for their responses (see \Cref{survey}). Next, they receive the aggregated responses and anonymized summaries of the rationales from other participants. The responses and rationales are then discussed in a workshop. After the workshop, participants have the opportunity to update their responses. These steps may be repeated until consensus is reached, but this is not strictly necessary. Finally, the responses are analyzed and a comprehensive report is prepared. The report includes areas of consensus, disagreement, and key insights from the expert panel, as well as a final assessment of the safety framework as a whole.

A key advantage of Delphi studies is that it leverages the insights of expert evaluators, which is particularly valuable in domains like AI Safety, where uncertainty is high, and best practices are still developing. Additionally, the interactive nature of the process encourages participants engage with new arguments~and diverging viewpoints during the grading process, which can lead to more thoughtful responses and consensus building.

One of the main disadvantages is that Delphi studies are time-consuming and require significant coordination effort (e.g. to schedule workshops), which can lead to low response rates. Participants might also update their responses in the direction of “leading” experts, rather than the strongest arguments \citep{aspinall2010}.

\hypertarget{audit}{%
\subsection{Audit}\label{audit}}

A final method for applying the grading rubric is through internal or external audits. This would involve two main components:

\textbf{Auditor selection.} Companies might commission a group of independent experts to evaluate their safety frameworks, such as academic institutions, civil society organizations, or an audit firm. Alternatively, they could assemble a group of individual experts, perhaps similar to red-teaming exercises.

\textbf{Audit process.} The grading rubric would serve as the audit standard. Auditors would be asked to evaluate each criterion on a scale from A to F. Importantly, these experts would be given access to non-public information to help them conduct their audits. To gather more information, they might also interview key personnel, review financial documents, and visit the developer's offices to perform or witness safety tests. Since they will have access to confidential information, they will likely have to sign non-disclosure agreements (NDAs).

The primary benefit of audits is that auditors can develop a more comprehensive understanding of a safety framework to inform their assessments. This is because auditors will typically have more time and access to more information, which can be a significant issue for other evaluation methods involving third parties. This is particularly useful for evaluating a framework's credibility (\Cref{credibility}), feasibility (\Cref{feasibility}), compliance (\Cref{compliance}), and empowerment (\Cref{empowerment}).

However, audits have some notable drawbacks. Firstly, they are significantly more time-consuming and potentially more costly than other evaluation methods. Secondly, the success of the audit relies on the developers' willingness to fully cooperate with the auditors and refrain from interfering with the process.

\hypertarget{limitations}{%
\section{Limitations}\label{limitations}}

In this section, we list six limitations of our proposed grading rubric. They should be kept in mind when grading safety frameworks and interpreting the results.

\textbf{Grading results might not translate into actionable recommendations.} While the evaluation criteria and indicators provide a useful tool for evaluating the quality of AI safety frameworks, they do not offer specific guidance on what commitments would be desirable (e.g. they do not contain any sample commitments). They provide a sense of \emph{what} to improve, but not \emph{how} to do so.

\textbf{The criteria are difficult to measure objectively.} Many evaluation criteria rely on abstract concepts such as robustness (see \Cref{robustness}), feasibility (\Cref{feasibility}), and transparency (\Cref{transparency}). These concepts are difficult to measure precisely and objectively (e.g. because it is difficult to quantify them). This is partly a feature rather than a bug of the design, as it may be difficult to specify precisely what a high quality framework contains. However, the lack of measurable indicators makes it challenging to evaluate some of the criteria. As a result, the evaluation process may heavily rely on qualitative assessments and subjective judgments, which can lead to inconsistencies and variations in the scores assigned by different graders. We encourage others to produce more easily assessed indicators as a means of grading safety frameworks.

\textbf{The criteria require evaluators to have AI safety expertise.} Some criteria, such as credibility (\Cref{credibility}) and robustness (\Cref{robustness}), require graders to have relevant AI safety and governance expertise. Since this expertise is scarce, it places limits on the number of people who could credibly assess the quality of frameworks. This seems unavoidable, but worth noting nonetheless.

\textbf{The evaluation criteria are unlikely to be exhaustive.} There may be additional factors inherent to what makes a safety framework ``good'' that are not adequately addressed by the proposed evaluation criteria. Safety frameworks are new and our current understanding of what makes a safety framework ``good'' is likely incomplete. As such, it is probable that there are factors and considerations missing which may prove to be critical in assessing the quality of safety frameworks in the future.

\textbf{It might be difficult to differentiate the six quality tiers.} The proposed grading system consists of six quality tiers, and the nuances that separate one tier from another may overly rely on subjective judgment, leading to potential inconsistencies and confusion in the scoring process. This issue is particularly relevant for the middle tiers, where the differences in quality may be less pronounced compared to the extreme ends of the scale. As a result, evaluators may find it difficult to assign scores with a high degree of precision, potentially limiting the usefulness of the grading system in providing a clear and reliable assessment of AI safety frameworks.

\textbf{The grading rubric does not weigh the evaluation criteria differently.} The criteria are unlikely to be equally important. For example, a framework that scores F on credibility (see \Cref{credibility}) and robustness (see \Cref{robustness}) but an A on all the other criteria would not be praiseworthy. This is part of why we do not recommend outputting a single overarching grade. However, even if the graders do not aggregate the scores, readers might intuitively do so, assigning equal weights to all criteria. As such, it might be important to produce a mapping between grades on all the criteria to an overall grade.\footnote{For example, you may want the categories to be multiples of each other. You may also want to put more weight on some things and disincentivize low scores across the board. A possible formula might therefore be: (weight 1 × Effectiveness) × (weight 2 × Adherence) × (weight 3 × Assurance).}

\section{Conclusion}\label{conclusion}

In this paper, we have proposed a grading rubric for AI safety frameworks. The rubric consists of seven grading criteria and 21 indicators that concretize the criteria. Each criterion can be graded on a scale from A (gold standard) to F (substandard). This grading rubric is our main contribution. We encourage governments, researchers, and civil society organizations to use the rubric to pass judgment on existing safety frameworks.

We wish to emphasize that developing an AI safety framework is extremely difficult. The first version of any safety framework will most likely be insufficient, and AI companies will need to continuously refine their frameworks in order to make them robust. However, they should not be the ones to decide whether their frameworks are adequate - they should not ``grade their own homework''. This should ultimately be the responsibility of governments with input from academia and civil society. We hope that our grading rubric can support this external scrutiny of AI safety frameworks.

\section*{Acknowledgements}\label{acknowledgements}

We are grateful for valuable feedback and suggestions from Alan Chan and Ben Garfinkel. All remaining errors are our own.

\bibliographystyle{abbrv}
\bibliography{ms}

\end{document}